# Control of the Optical Response of an Artificial Hybrid Nanosystem Due to the Plasmon-Exciton Plasmon Coupling Effect

Myong-Chol Ko,[1] Nam-Chol Kim,[1,2,*] Chol-Jong Jang,[1] Gwang-Jin Kim,[1] Zhong-Hua Hao,[2] Jian-Bo Li,[3] and Qu-Quan Wang[2,4,*]

[1]Department of Physics, **Kim Il Sung** University, Pyongyang, DPR of Korea
[2]School of Physics and Technology, Wuhan University, Wuhan 430072, China
[3]Institute of Mathematics and Physics, Central South University of Forestry and Technology, Changsha 410004, China
[4]The Institute for Advanced Studies, Wuhan University, Wuhan 430072, China
*ryongnam10@yahoo.com, qqwang@whu.edu.cn

**Abstract:** The optical response of an artificial hybrid molecule system composed of two metallic nanoparticles (MNPs) and a semiconductor quantum dot (SQD) is investigated theoretically due to the plasmon-exciton-plasmon coupling effects on the absorption properties of the hybrid nanosystem, which depends on the interaction between the induced dipole moments in the SQD and the MNPs, respectively. We show that the strong coupling of exciton and localized surface plasmons in such a hybrid molecules leads to appealing, tunable optical properties by adjusting the symmetry of the hybrid molecule nanosystem with controllable interparticle distances. We also address here the influence of the size of the MNPs and dielectric constant of the background medium on the optical absorption of the MNPs and SQD, respectively, which results in the interparticle Foster resonance energy transfer (FRET). Our results will open an avenue to deal with the surface-enhanced spectroscopies and potential application of the quantum information.

## 1. Introduction

The hybrid nanostructures based on metal nanoparticles (MNPs) and semiconductor quantum dots (SQDs) are the area of rapidly growing interest in a wide range of scientific fields, including physics [1], chemistry [2], medicine [3], and so on. Especially the interaction between excitons in SQDs and localized surface plasmons in MNPs leads to the considerable plexiton, which shows many interesting phenomena, such as plexciton quenching [4], chemical synthesis process [5], plasmon-assisted Forster energy transfer [6], induced exciton-plasmon-photon conversion [7], the sol–gel technique followed by thermal annealing [8], etc. The exciton of SQDs and surface plasmon field produced in MNPs interact via the long-range Coulomb force mechanism, the time of which depends strongly on the distances between SQDs and MNPs. Recently, Metal nanoparticles such as Ag nanoparticles which have been synthesized in physical vapor deposition method [9] show characteristic surface plasmon resonance (SPR) absorption in the visible region due to collective electron density oscillations at resonance frequencies. In the research of nanomaterials, much attention has been focused on the Fano effect in the energy absorption spectrum caused by the Fano interference between two competing optical pathways [10-15]. The Fano effect has already been experimentally observed in the plasmonic nanocrystal coupled with a polymer shell [11]. More interest in the interaction of the exciton and the plasmon provide the main key to the application of the quantum information processing such as single-photon transistors [16] or quantum switch [17]. However, much less is known on the role of the coupling effect between the SQD and the MNP. This is while the excited electronic states localized at SQD acting as an emitter placed close enough to the surface of plasmonic nanoparticle with the continuum of the electronic states of the metal. In previous studies it has shown the novel properties such as the Fano effect in energy absorption [13], bistability of the hybrid nanosystem [18], the generation of the plexciton [19], the exciton-induced transparency [20], which is composed of a SQD with two level and a MNP. Nowadays, it has rapidly growth of the research of optical properties of hybrid nanosystem with multi-particles [21-23] or multi-level quantum dot [24-27]. Furthermore, the energy absorption in a MNP-SQD-MNP hybrid system has been considered previously in Ref. [21], where they investigated Fano correlation effect on the basis of the cavity quantum electrodynamics and canonical transformation. By motivations, we discuss here theoretically the plasmon-exciton-plasmon coupling effects on the optical response, such as the energy absorption depending on the interparticle distances, which is addressed the interparticle Foster resonance energy transfer. We solve the optical master equation analytically via semi-classical approach and investigate the energy absorption spectrum of the nanoparticles.

2. **Theoretical Model and Solution**

We discuss a hybrid nanosystem composed of a single spherical SQD with radius r and two identical spherical MNPs of radii $a_1$, $a_2$, respectively, where the center-to-center interparticle distances with a MNP and a SQD are denoted $R_1$ and $R_2$, respectively. We consider a SQD with the vacuum ground state $|1\rangle$ and the excitonic state $|2\rangle$ in the presence of polarized external field $E = E_0 \cos(\omega t)$, where the direction of polarization is specified below as shown in Fig. 1. We treat the SQD as an effective two level quantum system with exciton energy $\hbar\omega_0$, transition dipole moment $\mu$, and dielectric constant $\varepsilon_s$ in the density matrix formalism. We treat the MNPs as classical spherical dielectric particles with dielectric function $\varepsilon_{M,1}(\omega)$ and $\varepsilon_{M,2}(\omega)$, respectively. We also regard the entire system to be embedded in a material of dielectric constant $\varepsilon_e$.

The Hamiltonian of the SQD with two levels can be written as follows [11]:

$$H_{SQD} = \hbar\omega_0 \hat{a}^+ \hat{a} - \mu E_{SQD}(\hat{a} + \hat{a}^+), \quad (1)$$

where $\hat{a}$ and $\hat{a}^+$ are the exciton annihilation and creation operator for SQD. $E_{SQD}$ is the electric field at the center of SQD that consists of the external field, $E$ and the induced fields produced by the polarization of the MNPs, $E_{M,1}$ and $E_{M,2}$. Thus, the total electric field felt on the SQD is $E_{SQD} = \frac{1}{\varepsilon_{effS}}(E + E_{M,1} + E_{M,2})$, where $\varepsilon_{effS} = \frac{2\varepsilon_e + \varepsilon_s}{3\varepsilon_e}$ is the screening factor of SQD and $E_{M,j} = \frac{1}{4\pi\varepsilon_e} \frac{s_\alpha P_{MNP,j}}{R_1^3}$ ($j = 1, 2$) is the field on SQD from the MNPs. $s_\alpha$ is equal to 2 (−1) when the interband dipole moment $\vec{\mu}$ is parallel(perpendicular) to the z (x, y) axis. The dipole $P_{MNP,j} = (4\pi\varepsilon_e)a^3 \gamma_j(\omega) E_{MNP,j}$ ($j = 1, 2$) comes from the charge induced on the surface of the jth MNP and depends on the total field due to the SQD, where $\gamma_j(\omega) = \frac{\varepsilon_{M,j}(\omega) - \varepsilon_e}{2\varepsilon_e + \varepsilon_{M,j}(\omega)}$ ($j = 1, 2$) is the dipole polarizability of the jth MNP. The total fields acting on the MNPs are just as follows;

$$E_{MNP,1} = E + \frac{s_\alpha P_{SQD}}{4\pi\varepsilon_e \varepsilon_{effS} R_1^3} + \frac{s_\alpha P_{MNP,2}}{4\pi\varepsilon_e \varepsilon_{effM,2}(R_1+R_2)^3},$$
$$E_{MNP,2} = E + \frac{s_\alpha P_{SQD}}{4\pi\varepsilon_e \varepsilon_{effS} R_2^3} + \frac{s_\alpha P_{MNP,1}}{4\pi\varepsilon_e \varepsilon_{effM,1}(R_1+R_2)^3}.$$
(2)

The dipole $P_{SQD}$ can be written via the off-diagonal elements of the density matrix as $P_{SQD} = \mu(\rho_{21} + \rho_{12})$. By using these relations we can define the parameters $\Omega$ and $G$, which present the normalized Rabi frequency associated with the external field and the field produced by the induced dipole moments $P_{MNP,j}$ of the $j$th MNP and the interaction between the polarized SQD and the MNPs, respectively.

$$\Omega = \frac{E_0 \mu}{2\hbar \varepsilon_{effS}} \left( 1 + \frac{\frac{s_\alpha a_1^3 \gamma_1}{R_1^3}\left(1 + \frac{s_\alpha a_2^3 \gamma_2}{\varepsilon_{effM,2}(R_1+R_2)^3}\right)}{1 - \frac{s_\alpha^2 a_1^3 a_2^3 \gamma_2 \gamma_1}{\varepsilon_{effM,1}\varepsilon_{effM,2}(R_1+R_2)^6}} + \frac{\frac{s_\alpha a_2^3 \gamma_2}{R_2^3}\left(1 + \frac{s_\alpha a_1^3 \gamma_1}{\varepsilon_{effM,1}(R_1+R_2)^3}\right)}{1 - \frac{s_\alpha^2 a_1^3 a_2^3 \gamma_2 \gamma_1}{\varepsilon_{effM,1}\varepsilon_{effM,2}(R_1+R_2)^6}} \right),$$

$$G = \frac{s_\alpha^2 \mu^2}{4\pi\varepsilon_e \hbar \varepsilon_{effS}^2} \left( \frac{\frac{a_1^3 \gamma_1}{R_1^3}\left(\frac{1}{R_1^3} + \frac{s_\alpha a_2^3 \gamma_2}{\varepsilon_{effM,2} R_2^3(R_1+R_2)^3}\right)}{1 - \frac{s_\alpha^2 a_1^3 a_2^3 \gamma_2 \gamma_1}{\varepsilon_{effM,1}\varepsilon_{effM,2}(R_1+R_2)^6}} + \frac{\frac{a_2^3 \gamma_2}{R_2^3}\left(\frac{1}{R_2^3} + \frac{s_\alpha a_1^3 \gamma_1}{\varepsilon_{effM,1} R_1^3(R_1+R_2)^3}\right)}{1 - \frac{s_\alpha^2 a_1^3 a_2^3 \gamma_2 \gamma_1}{\varepsilon_{effM,1}\varepsilon_{effM,2}(R_1+R_2)^6}} \right),$$

where $\varepsilon_{effM,j} = \frac{2\varepsilon_e + \varepsilon_{M,j}(\omega)}{3\varepsilon_e}$ ($j = 1, 2$).

We can present the simple expressions of the parameters $\Omega$ and $G$ by introducing the parameters, $K_i', K_i'', L_i', L_i'' (i = 1, 2)$, as follows;

$$\frac{\gamma_1\left(1+\dfrac{s_\alpha a_2^3 \gamma_2}{\varepsilon_{effM,2}(R_1+R_2)^3}\right)}{1-\dfrac{s_\alpha^2 a_1^3 a_2^3 \gamma_2 \gamma_1}{\varepsilon_{effM,1}\varepsilon_{effM,2}(R_1+R_2)^6}} = K'_1 + iK''_1, \qquad \frac{\gamma_1\left(\dfrac{1}{R_1^3}+\dfrac{s_\alpha a_2^3 \gamma_2}{\varepsilon_{effM,2}R_2^3(R_1+R_2)^3}\right)}{1-\dfrac{s_\alpha^2 a_1^3 a_2^3 \gamma_2 \gamma_1}{\varepsilon_{effM,1}\varepsilon_{effM,2}(R_1+R_2)^6}} = L'_1 + iL''_1,$$

$$\frac{\gamma_2\left(1+\dfrac{s_\alpha a_1^3 \gamma_1}{\varepsilon_{effM,1}(R_1+R_2)^3}\right)}{1-\dfrac{s_\alpha^2 a_1^3 a_2^3 \gamma_2 \gamma_1}{\varepsilon_{effM,1}\varepsilon_{effM,2}(R_1+R_2)^6}} = K'_2 + iK''_2, \qquad \frac{\gamma_2\left(\dfrac{1}{R_2^3}+\dfrac{s_\alpha a_1^3 \gamma_1}{\varepsilon_{effM,1}R_1^3(R_1+R_2)^3}\right)}{1-\dfrac{s_\alpha^2 a_1^3 a_2^3 \gamma_2 \gamma_1}{\varepsilon_{effM,1}\varepsilon_{effM,2}(R_1+R_2)^6}} = L'_2 + iL''_2.$$

We introduce the slowly varying quantities of the density matrix elements $\rho_{12} = \tilde{\rho}_{12} e^{i\omega_2 t}$ and $\rho_{21} = \tilde{\rho}_{21} e^{-i\omega_2 t}$ and can rewrite the Hamitonian of the SQD as follows;

$$\hat{H}_{SQD} = \hbar\omega_0 \hat{a}^+ \hat{a} - \hbar\left\{(\Omega + G\tilde{\rho}_{21})e^{-i\omega t} + (\Omega^* + G^* \tilde{\rho}_{12})e^{i\omega t}\right\}(\hat{a} + \hat{a}^+).$$

On the basis of the above relations, we can obtain the density matrix equations. From the solutions of the equations in the state limit, we can evaluate the energy absorptions of the nanoparticles. $Q_{SQD} = \hbar\omega_0 \rho_{22}/\tau_0$ is the energy absorption in the SQD. To calculate the energy absorbed by the MNP, we take the time average of the volume integral, $Q_{MNP} = \int j \cdot E dv$, where **j** is the current density and **E** is the electric field inside the MNPs. Thus, we obtain the energy absorption in MNPs as follows:

$$Q_{MNP,j} = 2\pi\varepsilon_B \omega a_j^3 \,\mathrm{Im}\!\left[\frac{\gamma_j}{\varepsilon^*_{effM,j}}\right](E_{Cj}^2 + E_{Sj}^2) \qquad (j=1,2),$$

where $E_{C2} = E_0 + \dfrac{s_\alpha \mu A}{2\pi\varepsilon_B \varepsilon_{effS} R_2^3} + \dfrac{s_\alpha a_1^3}{\varepsilon_{effM,1}(R_1+R_2)^3}\left(E_0 K'_1 + \dfrac{s_\alpha \mu A}{2\pi\varepsilon_B \varepsilon_{effS}} L'_1 + \dfrac{s_\alpha \mu B}{2\pi\varepsilon_B \varepsilon_{effS}} L''_1\right)$ and

$E_{S2} = \dfrac{s_\alpha \mu B}{2\pi\varepsilon_B \varepsilon_{effS} R_2^3} - \dfrac{s_\alpha a_1^3}{\varepsilon_{effM,1}(R_1+R_2)^3}\left(E_0 K''_1 - \dfrac{s_\alpha \mu B}{2\pi\varepsilon_B \varepsilon_{effS}} L'_1 + \dfrac{s_\alpha \mu A}{2\pi\varepsilon_B \varepsilon_{effS}} L''_1\right).$

## 3. Numerical Results and Theoretical Analysis

In the following we consider the exciton-plasmon coupling effects on the energy absorption of the nanoparticles through formation of plexciton in the strong field with its intensity, $I = 1\text{kW}/\text{Cm}^2$, numerically, where the SQD of radius $a = 0.65nm$ has two-level structure with the dielectric constant, $\varepsilon_S = 6.0$ and the MNPs are gold nanoparticles with the bulk dielectric constant, $\varepsilon_M(\omega)$ taken from Ref. [27]. The bare exciton frequency, $\omega_0$, is set to be 2.5 eV, close to the surface plasmon resonance of the Au MNP.

First of all, we discuss the energy absorption spectra in MNP and SQD versus the detuning, $\omega - \omega_0$, when the interparticle distance between one of two MNPs and the SQD is fixed with 13nm and the other interparticle distance can be different (Fig. 2). In Fig. 2 (a) the height of peak of the energy absorption in MNP gets low as the interparticle distance gets long, which means the FRET from SQD to MNP gets small. This is why the exciton-plasmon coupling effect depends on the interparticle distance. In particular, when the interparticle distance is 7nm, double peaks appear in the energy absorption spectrum, which shows the optical bistability of the MNP. We can also find the usual linear Fano effect which the energy absorption becomes zero for a particular frequency due to the interference effect. And in the other cases the energy absorption spectra have the asymmetrical Fano shape. From Fig. 2(b) you can see the asymmetry of the energy absorption in the SQD but the top of its peak is placed at the resonant frequency with the natural frequency of the SQD. The asymmetry gets strong when interparticle distances is big different.

Fig. 3 shows the ratio of the energy absorption in MNPs versus the detuning, $\omega - \omega_0$, when the interparticle distances are variant, that is, the hybrid system is asymmetry or symmetry. In Fig. 3(a), a narrow peak appears at the positive detuning, which shows the energy absorption in the MNP with interparticle distance of 13nm is much bigger than one in the other MNP with interparticle distance of 7nm in certain range of the frequency. As shown in Fig. 3(b)~ (d), the ratio of the energy absorption has the various line shape such as asymmetric Fano line shape, Fano line shape, straight line and double peaks line. From these figures we can find the energy absorption rate in MNPs according to the interparticle distance intuitively. In the other words, the FRET from SQD to the MNP gets large because the exciton-plasmon coupling gets strong when the interparticle distance between the SQD and MNP gets small.

We also investigate the influence of the size of the MNP on the energy absorption in MNPs, where the MNPs are equally spaced from the SQD with the interparticle distance of 13nm. As you can see from Fig. 4, the absorption spectra has the linear Fano effect line shape and the energy absorption in the MNP with the radius of 5nm is remarkably bigger than one in the MNP with the radius of 3nm, which shows that the property of the plexciton generated from the dipole-dipole interaction between the SQD and the MNP is related with not only the interparticle distance, but also the size of nanoparticles. Thus, we can find the intensity of the localized field due to the charge induced on the surface of the MNPs gets strong when the size of the MNP becomes bigger.

Next, we consider the influence of the background medium with the different dielectric constants on the absorption spectra, where the MNPs have the same size and are equally spaced from the SQD. The FRET from SQD to every MNP is equal to each other since the interparticle distances between the SQD and the MNPs are the same, which means the same plexciton is generated from the interaction between the SQD and the MNP. By the reason, Fig. 5 plots the energy absorption spectra of a MNP in the different background medium. When the dielectric constant of the background medium gets big, the inherent localized fields of the nanoparticles become strong, which leads to the strengthened interference effect totally. Therefore, the peak of the absorption spectrum with the dielectric constant of background medium, 3, is higher than the other cases.

Finally, we address the energy absorption in the SQD versus the ratio of the interparticle distances when the frequency of the external field is resonant with the natural frequency of the SQD. Fig. 6 shows the variation of the peak of the energy absorption when sizes of the MNPs are the same or not. The maximum value gets small when the ratio of the interparticle distances gets big, which shows the FRET form the SQD to the MNP becomes large. Especially, when the size of the MNP is 5nm (dashed line), the height of the peak of the absorption spectrum becomes low rapidly.

## 4. Conclusion

In conclusion, we have studied theoretically the optical response of an artificial hybrid molecule system composed of two metallic nanoparticles (MNPs) and semiconductor quantum dot (SQD) due to the plasmon-exciton-plasmon coupling effects on the absorption properties of the hybrid nanosystem. It is shown that the energy absorption spectra depend on the interaction with the induced dipole moments in the SQD and the MNPs, respectively. We show that the strong coupling exciton and localized surface plasmons in such a hybrid molecules provides us interesting Fano interference process by adjusting the symmetry of the hybrid molecule nanosystem with controllable interparticle distances. We also find here the influence of the size of the MNPs, the ratio of the interparticle distances and dielectric constant of the background medium on the optical absorption of the MNPs and SQD, respectively. Thus, the hybrid nanosystem discussed in this paper share a common segment of optical pathway, where there is the FRET which is a central topic in quantum optics. Our results will open a way to deal with the plasmon-enhanced devices and the photovoltaic device with the potential application of the quantum information.

**Acknowledgments.** This work was supported by the National Program on Key Science Research of DPR of Korea (Grant No. 131-00). This work was also supported by the National Program on Key Science Research of China (2011CB922201) and the NSFC (11174229, 11204221, 11374236 and 11404410) and the Foundation of Talent Introduction of Central South University of Forestry and Technology (104-0260).

Figure Caption

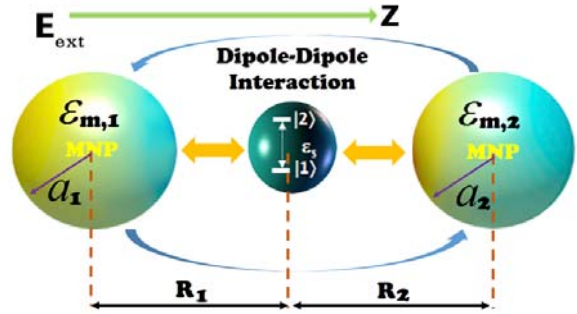

**Fig. 1** (Color online). Schematic diagram of the MNP-SQD-MNP hybrid nanostructure. $a_1$ and $a_2$ are the radii of MNPs, respectively. $R_1$ and $R_2$ are the center-to-center distances between SQD and MNPs, respectively. $\varepsilon_e$, $\varepsilon_s$ and $\varepsilon_m$ are the dielectric constants of the background medium, SQD and MNP, respectively.

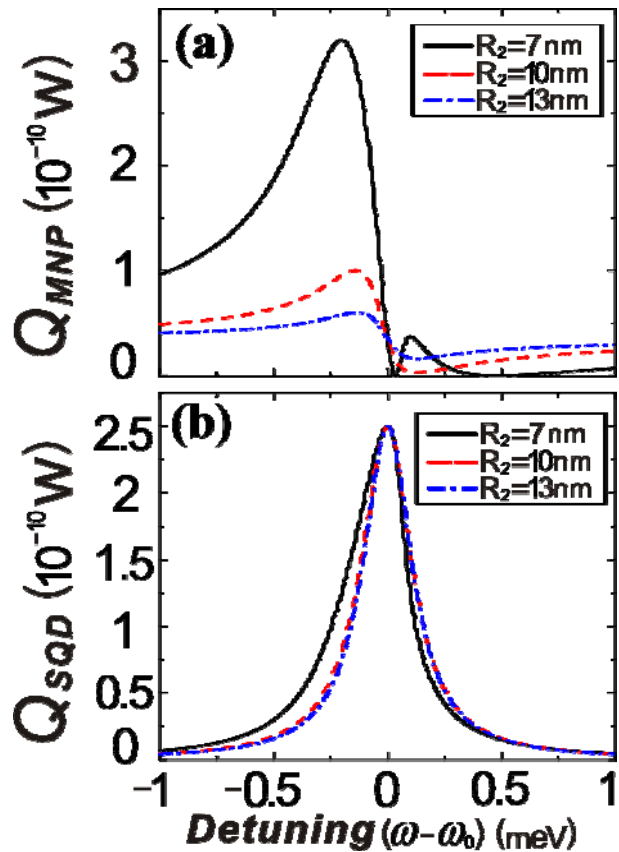

**Fig. 2** (Color online). The energy absorption spectra versus the detuning, $\omega - \omega_0$, when one of the interparticle distances are 7nm(solid line), 10nm(dashed line), 13nm(dash-dotted line),

respectively. (a) in MNP (b) in SQD. Here, we set $\varepsilon_e =1$, $a_1= a_2=3$nm, $r=0.65$nm, $R_1=13$nm, $S_\alpha=2$ and I=1kW/cm$^2$.

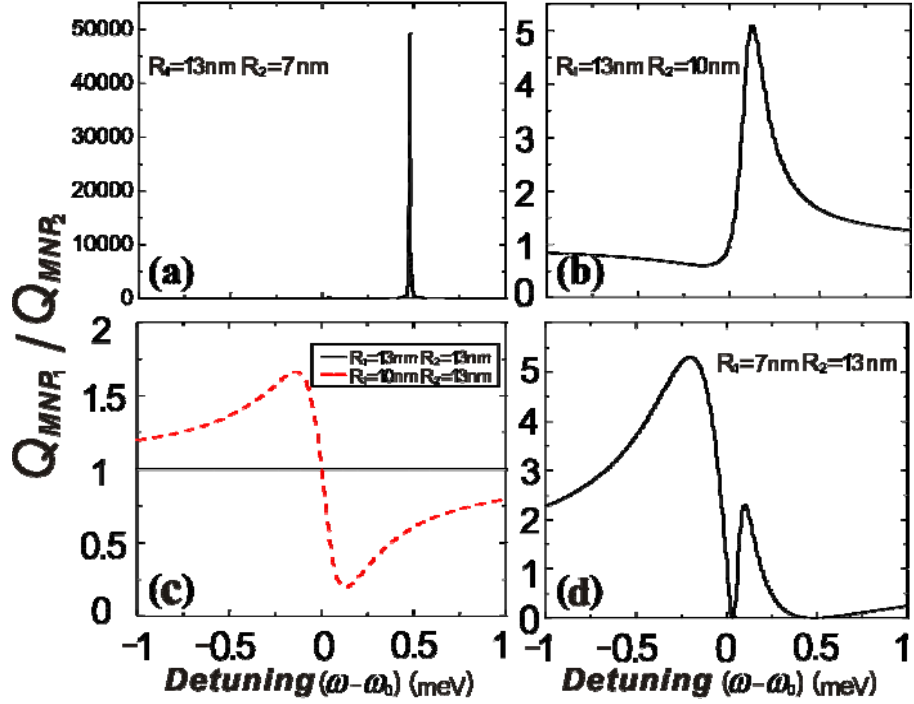

**Fig. 3** (Color online). The ratio of the energy absorption in MNPs versus the detuning, $\omega-\omega_0$, when the interparticle distances are different. (a) $R_1=13$nm, $R_2=7$nm, (b) $R_1=13$nm, $R_2=10$nm, (c) $R_1=13$nm, $R_2=13$nm and $R_1=10$nm, $R_2=13$nm, (d) $R_1=7$nm, $R_2=13$nm. Here, we set $\varepsilon_e =1$, $a_1= a_2=3$nm, $r=0.65$nm, $S_\alpha=2$ and I=1kW/cm$^2$.

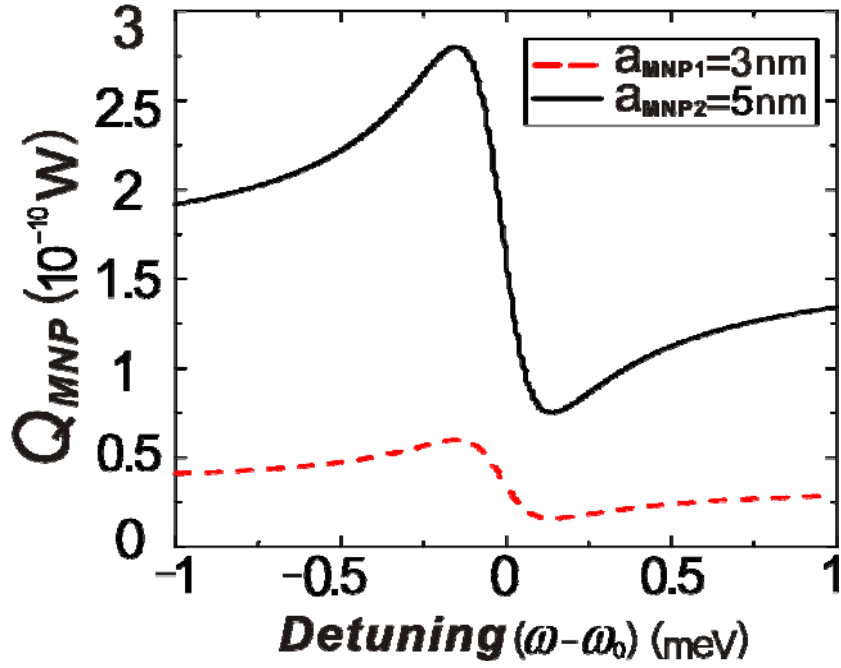

**Fig. 4** (Color online). Energy absorption spectra for different sizes of the MNPs versus the detuning, $\omega-\omega_0$. Here, we set $\varepsilon_e=1$, $a_1$=3nm, $a_2$=5nm, $r$=0.65nm, $R_1$=$R_2$=13nm, $S_\alpha$=2 and I=1kW/cm$^2$.

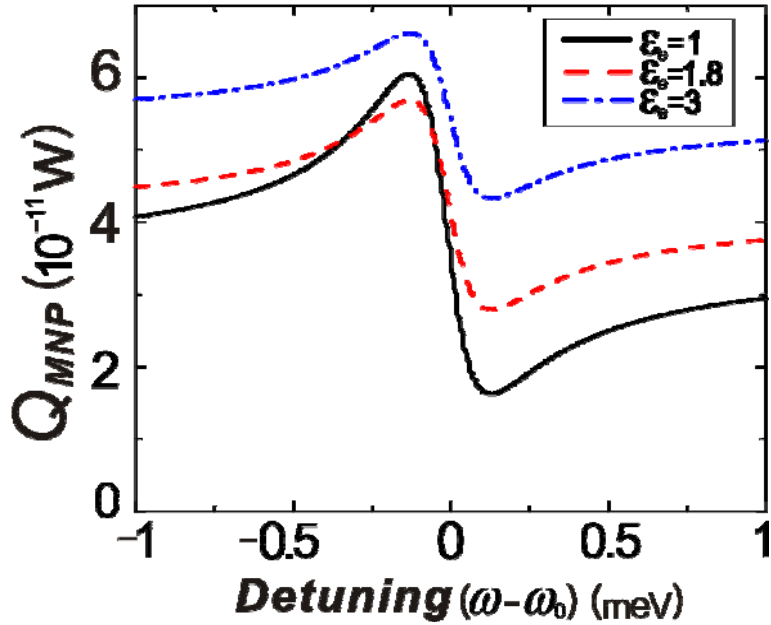

**Fig. 5** (Color online). The influence of the background medium on the absorption spectra in MNP versus the detuning, $\omega-\omega_0$. Here, we set $a_1$=3nm, $a_2$=3nm, $r$=0.65nm, $R_1$=$R_2$=13nm, $S_\alpha$=2 and I=1kW/cm$^2$.

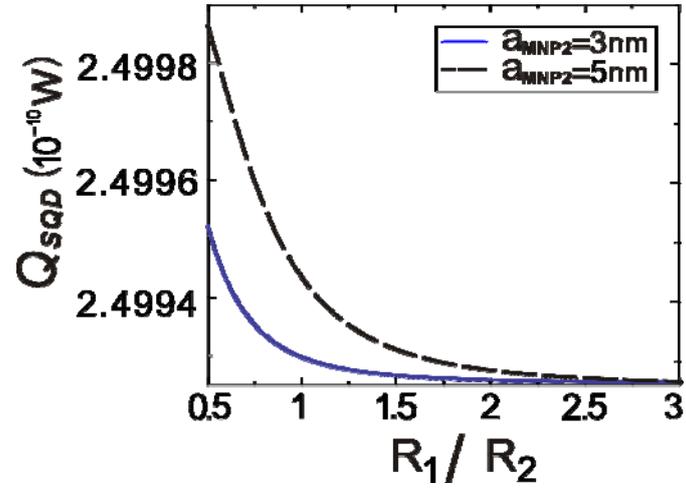

**Fig. 6** (Color online). Energy absorption spectra in the SQD versus the ratio of the interparticle distances. Here, we set $\varepsilon_e = 1$, $a_1$=3nm, $r$=0.65nm, $S_\alpha$=2, $\omega = 2.5$eV and I=1kW/cm$^2$.